\documentclass{article}

\usepackage{hyperref}
\usepackage[left=2.0cm,right=2.0cm,top=2.5cm,bottom=2.5cm]{geometry}
\usepackage{graphicx}
\usepackage{amssymb,amsmath,amscd}
\usepackage{mathtools, color}
\usepackage{cleveref}
\usepackage{multirow}

\usepackage{xcolor}
\usepackage{dcolumn}
\usepackage{bm}

\usepackage[utf8]{inputenc}
\usepackage[T1]{fontenc}
\usepackage{mathptmx}
\usepackage{etoolbox}

\newcommand{\eqN}[1]{(\ref{#1})} 

\title{Restoring the reciprocity invariance in nonlinear systems with broken mirror symmetry}
\author{Andrus Giraldo\footnotemark[1] 
\and Behrooz Yousefzadeh\footnotemark[2]}

\begin{document}

\renewcommand{\thefootnote}{\fnsymbol{footnote}}
\footnotetext[1]{School of Computational Sciences, Korea Institute for Advanced Study, Seoul 02455, Korea

  (\href{mailto:agiraldo@kias.re.kr}{agiraldo@kias.re.kr})}
\footnotetext[2]{Department of Mechanical, Industrial \& Aerospace Engineering, Concordia University, Montreal, QC, H3G1M8, Canada

  (\href{behrooz.yousefzadeh@concordia.ca}{behrooz.yousefzadeh@concordia.ca})}
\renewcommand{\thefootnote}{\arabic{footnote}}

\date{\today}

\maketitle

\begin{abstract}
Circumventing the reciprocity invariance has posed an interesting challenge in the design of modern devices for wave engineering. In passive devices, operating the device in the nonlinear response regime is a common means for realizing nonreciprocity. Because mirror-symmetric systems are trivially reciprocal, breaking the mirror symmetry is a necessary requirement for nonreciprocal dynamics to exist in nonlinear systems. However, the response of an asymmetric nonlinear system is not necessarily nonreciprocal. In this work, we report on the existence of stable, steady-state nonlinear reciprocal dynamics in coupled asymmetric systems subject to external harmonic excitation. We restore reciprocity in the asymmetric system by tuning two symmetry-breaking parameters simultaneously. We identify response regimes in the vicinity of the primary resonances of the system where the steady-state left-to-right transmission characteristics are identical to the right-to-left characteristics in terms of frequency, amplitude and phase. We interpret these regimes of reciprocal dynamics in the context of phase nonreciprocity, wherein incident waves undergo a nonreciprocal phase shift depending on their direction of travel. 
We hope these findings help design devices with new functionalities for controlling and steering of elastic waves.

\end{abstract}

Reciprocity is an important invariance property of wave propagation in materials, a property fundamental to electronics~\cite{Reciprocity_electronics}, electromagnetics~\cite{Reciprocity_electromagnetics}, acoustics\cite{tenWolde,Fahy_Acta,BY_NRM}, elasticity~\cite{Achenbach_book,BY_NRM} and fluid dynamics~\cite{Reciprocity_Godin,Reciprocity_Stone}. Reciprocity invariance is a form of symmetry in wave phenomena that ensures waves travel identically across a transmission channel independent of the direction of propagation. In other words, the ``left-to-right'' transmission characteristics between two points (frequency, amplitude and phase) are identical to the ``right-to-left'' characteristics. 
Devices that can operate beyond this invariance property enable control of wave propagation in a directional manner: asymmetric propagation along opposite directions. Certain well-established communication devices such as isolators and circulators rely on nonreciprocity for their operation~\cite{Reciprocity_electronics}. 

While designing against the reciprocity invariance has a long history in electronics and electromagnetics, developing nonreciprocal devices is also experiencing a surge of interest in acoustics and elasticity~\cite{BY_NRM,NonrecipAcoustics}.  
In the absence of the action of external biases such as moving boundaries or fluid flow, the reciprocity invariance holds in linear, time-invariant materials with symmetric constitutive relations. Our focus here is on passive (time-invariant) devices, for which nonreciprocal transmission can be realized by operating the device in the nonlinear response regime.
Mirror-symmetric transmission channels are trivially reciprocal owing to the invariance caused by their symmetry. Thus, breaking the mirror symmetry of a nonlinear system is necessary for enabling a nonreciprocal response~\cite{LepriCasati,nonlinearity_AluI}. While asymmetry is a necessary condition for nonreciprocal dynamics in nonlinear systems, it is not a sufficient condition~\cite{BY_NRM,volterraJSV,ren2019}. It is, therefore, possible to realize reciprocal dynamics in a nonlinear system with broken mirror symmetry. 

In this work, we report on the existence of stable, steady-state nonlinear reciprocal dynamics in coupled asymmetric systems. We identify response regimes in which the forward (left-to-right) transmission characteristics -- in terms of frequency, amplitude and phase -- are identical to the backward (right-to-left) transmission characteristics in response to external harmonic excitation. We relate the reciprocal dynamics to the concept of \emph{phase nonreciprocity}: response regimes where nonreciprocity manifests itself as a difference in phase, but not amplitude, of the forward and backward transmission paths~\cite{BY_JCD}. Within this context, a reciprocal response corresponds to a zero difference between the phase shifts in the forward and backward configurations.  

\begin{figure*}
  	 \centering
    	 \includegraphics{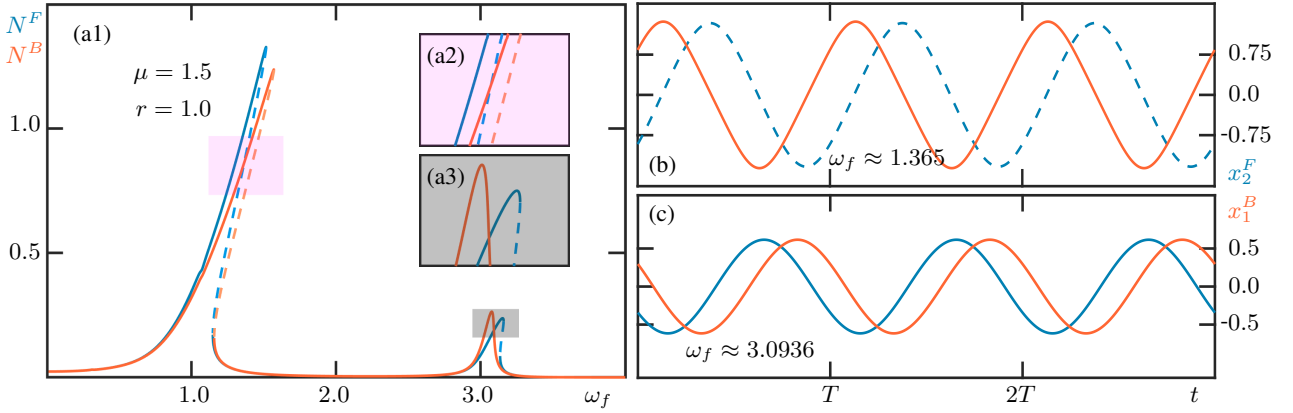}
   	 \caption{Phase nonreciprocity in the coupled nonlinear system. Panel~(a1) shows the output norms for the forward configuration, blue curve ($N^F$), and the backward configuration, orange curve ($N^B$), as a function of the forcing frequency, $\omega_f$.   The insets~(a2) and (a3) show magnifications of the respective colored regions in panel~(a1) near the points where the curves $N^F$ and $N^B$ intersect. The temporal traces for the forward output displacement $x^F_2$, blue curve, and the backward output displacement $x^B_1$, orange curve, are shown in panels~(b) and (c) at the values of $\omega_f$ (indicated in each panel) where $N^F=N^B$ in panels~(a2) and (a3). In all panels, solid lines represent stable responses, and dashed lines represent unstable responses. Other parameters which are not indicated in the figure are $(\zeta,k_c,k_N,P)=(0.05, 5, 1,0.5)$.} 
\label{fig1}
\end{figure*}

We consider the archetypal framework of two coupled oscillators to demonstrate the reciprocal dynamics and nonreciprocal phase shifts in nonlinear asymmetric systems. Coupled oscillators, in addition to providing a straightforward discrete platform for investigating nonreciprocity, describe the dynamics of more complicated systems ranging from coupled elastic and optical waveguides~\cite{waveguide} to two-port electromagnetic resonators~\cite{nonlinearity_AluI,garbin}, optical frequency combs~\cite{comb} and hopping bosons~\cite{AndrusBoseHubbard}. The governing equations for coupled nonlinear oscillators that we consider are 
\begin{equation}
	\label{eq:EOM}
	\begin{aligned}
	\ddot{{x}}_1+ 2\zeta \dot{{x}}_1 + {x}_1 + k_c (x_1-x_2) + k_N x_1^3 &= F_1 \cos(\omega_f t), \\
	\mu \ddot{{x}}_2+ 2\zeta \dot{{x}}_2 + r {x}_2 + k_c (x_2-x_1) + \alpha k_N x_2^3 &= F_2 \cos(\omega_f t)	,	
    \end{aligned}
\end{equation}
where $k_c$ represents the strength of coupling between the two systems and $k_N$ the strength of nonlinearity. The mirror symmetry is controlled by parameters $\mu$ and $r$ in the linear response regime and by $\alpha$ at higher amplitudes of motion. In a mechanical realization of the system, $\mu$, $r$, and $\alpha$ correspond respectively to the ratios of the masses, grounding springs, and nonlinear elasticity of the individual (uncoupled) oscillators. Energy losses are modeled by a linear viscous damping mechanism with $\zeta=0.05$. We present our main findings for strong coupling ($k_c=5$) and hardening nonlinearity ($k_N=1>0$) as a representative example. The existence of stable reciprocal dynamics, however, is not specific to these parameter values.

The two configurations needed to test for reciprocity invariance are (i) the forward or left-to-right (L--R) configuration, where $F_1=P$, $F_2=0$ and $x_2^F(t)$ is the output displacement; (ii) the backward or right-to-left (R--L) configuration, where $F_1=0$, $F_2=P$ and $x_1^B(t)$ is the output displacement. Reciprocity holds if and only if $x_2^F(t)=x_1^B(t)$. 

To quantify the response of the system, we define the forward and backward output norms, $N^F$ and $N^B$, where $N^F=(1/T)\int_0^T x_2^F(t)^2dt$ and $N^B=(1/T)\int_0^T x_1^B(t)^2dt$ are defined over a forcing period $T=2\pi/\omega_f$. 
To quantify the degree of nonreciprocity, we define the following norm: 
\begin{equation}
    \label{Rnorm}
    R = \frac{1}{T} \int_0^T \left(x_2^F(t)-x_1^B(t)\right)^2dt 
\end{equation} 
The response is reciprocal if and only if $R=0$.
We use {\color{black} and develop} numerical  continuation techniques  through the software package {\sc auto}~\cite{auto_old,auto} to compute the steady-state response of Eq.~\eqN{eq:EOM} as a family of periodic orbits~\cite{Seb_ch1,Seb_handbook} {\color{black} that satisfies a suitable two-point boundary value problem}. This formulation allows for simultaneous computation of all the relevant parameters and norms, including the nonreciprocal phase shift, thus eliminating the need for lengthy post-processing of data.

We present the reciprocal dynamics of the system in the context of phase nonreciprocity~\cite{BY_JCD}. Fig.~\ref{fig1}(a1) shows the frequency response curve of the asymmetric system with $\mu=1.5$. Because of asymmetry ($\mu\ne1$), the response near each primary (one-to-one) resonance is markedly nonreciprocal. The insets (a2) and (a3) highlight the portions of the frequency response curves where the two output norms are equal, $N^F=N^B$. Panels (b) and (c) show the corresponding output displacements for the first (a2) and second (a3) modes~\cite{modes}. Reciprocity invariance does not hold at these two values of $\omega_f$ because the output displacements have different phases. In other words, there is a nonreciprocal phase shift, $\Delta\phi$, between $x_2^F(t)$ and $x_1^B(t)$ at these frequencies. A detailed discussion of nonreciprocal phase shifts in nonlinear coupled oscillators can be found elsewhere~\cite{BY_JCD}.

Keeping the output norms equal, $N^F=N^B$, a reciprocal response is obtained if the nonreciprocal phase shift can be set to zero, $\Delta\phi=0$. The symmetry-breaking parameter $\mu$ is already used to break the reciprocity invariance. Therefore, a second symmetry-breaking parameter, $r$ or $\alpha$, is required to realize $\Delta\phi=0$ and restore the invariance condition. The two symmetry-breaking parameters would be, in effect, balancing each other to maintain reciprocity. A similar discussion can be found in the context of  symmetry-breaking bifurcations~\cite{imperfectSIAM,imperfect}, for example in nonlinear optical devices~\cite{garbin}.

\begin{figure}
  \centering
     \includegraphics[width=10cm]{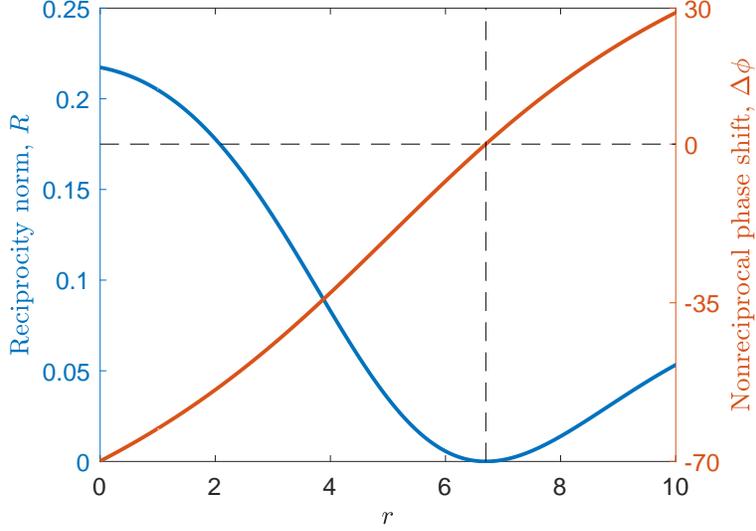}
   	 \caption{Evolution of the locus of phase nonreciprocity ($N^F=N^B$) as a function of the symmetry-breaking parameter $r$ for the second (out-of-phase) mode. The zeros of $R$ and $\Delta\phi$ coincide near $r=6.70$. Other parameters in Eq.~\eqN{eq:EOM} which are not indicated in the figure are $(\mu, \zeta,k_c,k_N,P)=(1.5, 0.05, 5, 1,0.5)$.} 
\label{fig2}
\end{figure}

\begin{figure*}
  	 \centering
    	 \includegraphics{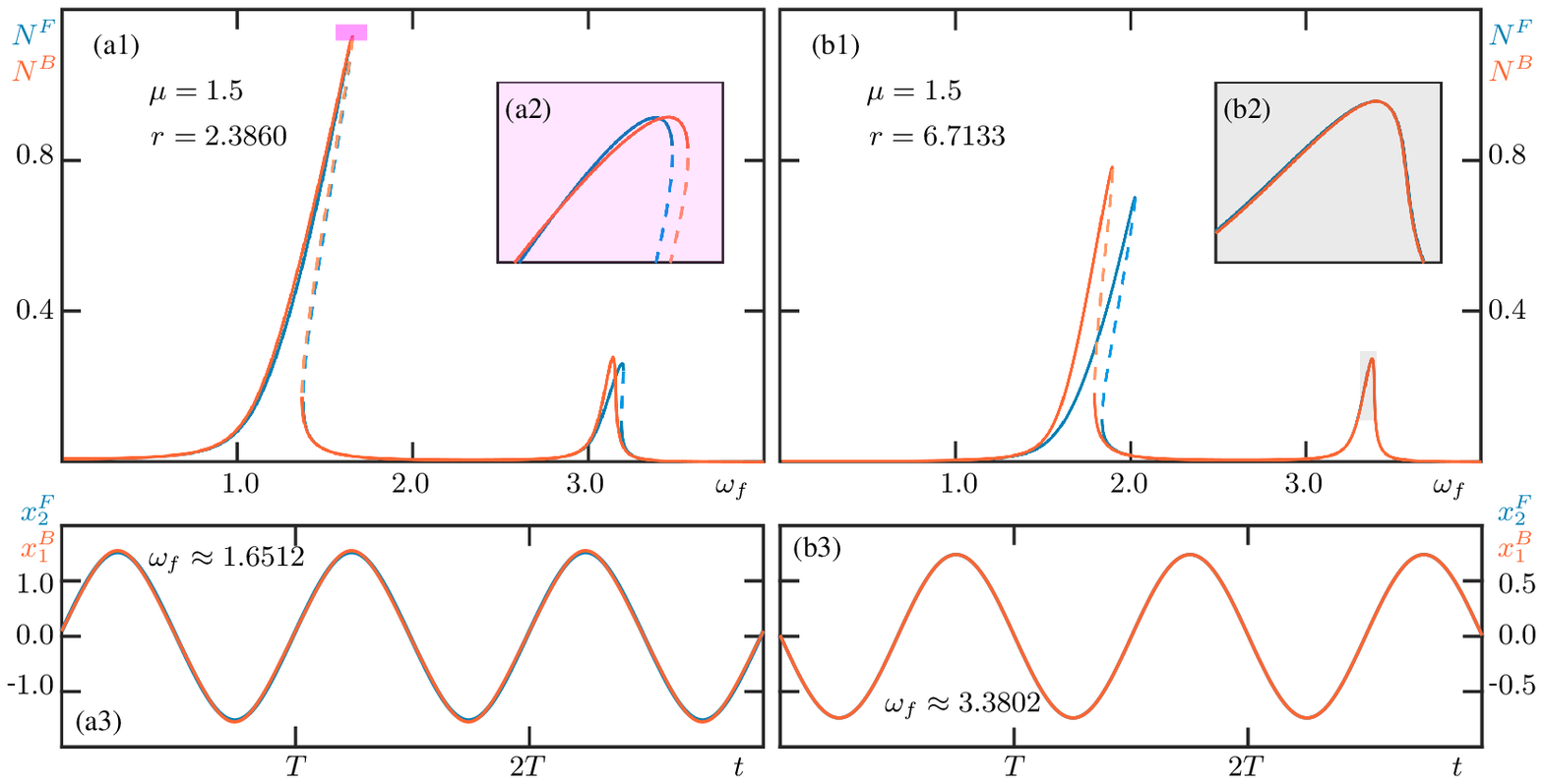}
   	 \caption{Reciprocal dynamics in the system with broken mirror symmetry. Panel~(a1) and (b1) show the output norms for the forward configuration, blue curve ($N^F$), and the backward configuration, orange curve ($N^B$), as a function of the forcing frequency, $\omega_f$, for $\mu=1.5$ and two values of $r$.   The insets~(a2) and (b2) show magnifications of the respective colored regions in panels~(a1) and (a2), respectively, near the intersection points where reciprocity is achieved ($R\approx0$). The reciprocal response for the forward output displacement $x^F_2$, blue curve, and the backward output displacement $x^B_1$, orange curve, are shown in panels~(a3) and (b3). In all panels, solid curves represent stable response and dashed curves represent unstable response; 
     $R=3.287\cdot10^{-4}$ in (a3) and $R=6.172\cdot10^{-8}$ in (b3). Other parameters which are not indicated in the figure are $(\zeta,k_c,k_N,P)=(0.05, 5, 1,0.5)$. 
     } 
\label{fig3}
\end{figure*}

Fig.~\ref{fig2} shows the evolution of the locus of phase nonreciprocity ($N^F=N^B$) as a function of the symmetry-breaking parameter $r$ for the second mode at $P=0.5$. The reciprocity norm vanishes ($R=0$) near $r^\star\approx6.70$. This point coincides with $\Delta\phi=0$, as anticipated. The reciprocity invariance is therefore restored with two symmetry-breaking parameters $(\mu,r)\approx(1.5,6.70)$ at this particular operating point.

Fig.~\ref{fig3} shows the frequency response curves of the system for $\mu=1.5$ at $P=0.5$. The second symmetry-breaking parameter $r$ is chosen to realize reciprocal dynamics for the first and second modes; see panels (a1) and (b1), respectively. The insets (a2) and (b2) highlight the portions of the frequency response curves where $R=0$ at the intersection of the two curves. Panels (a3) and (b3) show the corresponding output displacements of the system, which are obtained by direct numerical integration of the governing equations. The steady-state response for both the forward and backward configurations is stable for the chosen parameters. 

We note in panels (a1) and (b1) of Fig.~\ref{fig3} that the response of the system is nonreciprocal except near the specific points highlighted in panels (a2) and (b2). At these points, the balance between the two symmetry-breaking parameters is such that reciprocity is restored (b3) or approximately restored (a3). A significant practical feature of the reciprocal dynamics highlighted in Fig.~\ref{fig3} is that they occur within the primary resonance regimes of the response. Therefore, the predicted reciprocal dynamics can be observed in an experiment; monitoring the nonreciprocal phase shift may prove very useful for this purpose. {\color{black} For the cases shown, along the first mode in panel~(a1) and the second mode in panel~(b1), approximate reciprocity ($R \approx 0$) is also attained along each of the resonances, even though our computational techniques only pertain to the intersection points between $N^F$ and $N^B$.} A very small value of $R$ can also be obtained between the two peaks; e.g., for $2.0<\omega_f<2.4$ in Fig.~\ref{fig3}(a1). However, the amplitudes of motion in these regions (anti-resonances) are too small to be measured reliably.  

\begin{figure}
  	 \centering
    	 \includegraphics{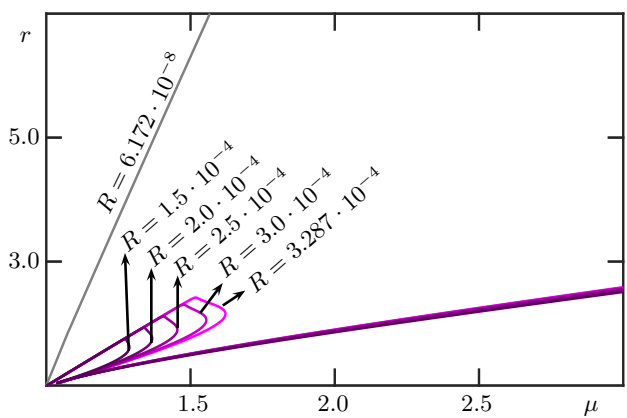}
   	 \caption{Loci of approximate reciprocity $R=\epsilon$ in the $(\mu,r)$-parameter plane. Purple lines corresponds to the locus of reciprocity near the first primary resonance, where darker shades represents lower values of $\epsilon$. The grey line corresponds to the locus of reciprocity near the second primary resonance. Solid lines represent stable solutions and dashed lines represent unstable solutions. Other parameters which are not indicated in the figure are $(\zeta,k_c,k_N,P)=(0.05, 5, 1,0.5)$.} 
\label{fig:LocusNonReciprocity}
\end{figure}

A note on the computation of the locus of $R=0$ is in order here. Because $R$ is a non-negative quantity, it is numerically difficult to ascertain when it becomes zero. The simultaneous computation of the nonreciprocal phase shift facilitates finding points at which $R=0$, as indicated in Fig.~\ref{fig2}. The computation of the nonreciprocal phase shift itself is straightforward when the response is harmonic, which is the case for the results shown in Fig.~\ref{fig2}. To quantify this observation, the \emph{harmonicity} of the output for the forward and backward configurations were computed as $H^{F,B}=(a^{F,B})^2/2N^{F,B}$, whereby a harmonicity of 1 indicates a purely harmonic response. A minimum harmonicity of $0.999~998$ is computed for the response in Fig.~\ref{fig2}. It is conceivable that the harmonicity of the response is compromised at higher values of $P$ or $k_N$. In this case, it remains straightforward to compute the locus of $R=\epsilon$ with $\epsilon\ll 1$ to approximate the locus of reciprocity. 

Having demonstrated the possibility of realizing stable reciprocal dynamic response near the primary resonances of the system at a particular set of parameters, we now turn our attention to finding other sets of parameters for which {\color{black} approximate} reciprocal response ({\color{black} $R\approx0$}) can be achieved. 
Fig.~\ref{fig:LocusNonReciprocity} shows the loci of reciprocal response in the $(\mu,r)$-parameter space for the two primary resonances of the system. Using a harmonic approximation of the response, we show in the Appendix that the locus of $R=0$ satisfies the following relation when $\alpha=1$:
\begin{equation}
    \label{eq:linRelation}
    (r-1)-(\mu-1)\omega^2_f=0
\end{equation}
where $\omega_f$ is the forcing frequency at which $R=0$.

\begin{figure*}
  	 \centering
    	 \includegraphics{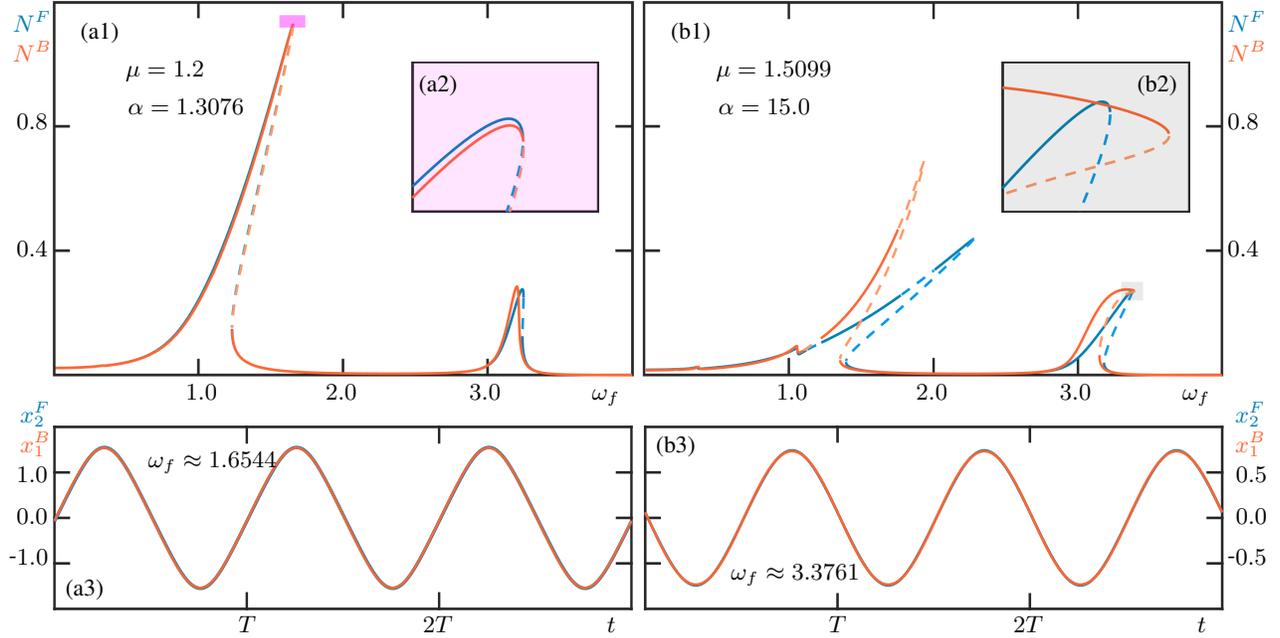}
   	 \caption{Reciprocal dynamics in the system with broken mirror symmetry. Panel~(a1) and (b1) show the output norms for the forward configuration, blue curve ($N^F$), and the backward configuration, orange curve ($N^B$), as a function of the forcing frequency, $\omega_f$, for two different sets of values for $\mu$ and $\alpha$.  The insets~(a2) and (b2) show magnifications of the respective colored regions in panels~(a1) and (a2), respectively, near the intersection points where reciprocity is achieved ($R\approx0$). The reciprocal response for the forward output displacement $x^F_2$, blue curve, and the backward output displacement $x^B_1$, orange curve, are shown in panels~(a3) and (b3). In all panels, solid curves represent stable response and dashed curves represent unstable response; solutions in panels~(a3) and (b3) are stable. Other parameters which are not indicated in the figure are $(r,\zeta,k_c,k_N,P)=(1,0.05, 5, 1,0.5)$.} 
\label{fig5}
\end{figure*}

Fig.~\ref{fig:LocusNonReciprocity} shows that the computed locus of reciprocal response is very closely approximated by a line for the second mode (corresponding to the right column in Fig.~\ref{fig3}). The relation between $r$ and $\mu$ is linear in this case because $\omega_f$ remains relatively unchanged along the locus of $R\approx0$; \emph{i.e.}, $\omega_f \in (3.30,3.3766)$. The slope of the corresponding line in Fig.~\ref{fig:LocusNonReciprocity} is approximately $11.2(=3.35^2)$, as predicted by Eq.~\eqN{eq:linRelation}. Further computations (not reported) suggest that the locus persists in the same fashion up to $\mu=4$ for this mode.

The locus of reciprocal dynamics in Fig.~\ref{fig:LocusNonReciprocity} looks different for the first mode (corresponding to the left column in Fig.~\ref{fig3}). Starting from the trivial (mirror-symmetric) case at $(\mu,r)=(1,1)$, the locus starts linearly with $\omega_f \in [1.6512,1.6553)$ and a corresponding slope of $2.7(=1.65^2)$. The response along this segment of the locus is near the primary resonance of the system; see Fig.~\ref{fig3}(a1). As we continue along the locus, however, there is a sharp bend toward the mirror-symmetric configuration, which is accompanied by a decrease in the forcing frequency (down to $\omega_f=1.1480$). This transition point depends on the numerical tolerance $\varepsilon$ used for computing $R=\varepsilon\approx0$; for tighter tolerances (smaller $\varepsilon$), the transition occurs closer to the mirror-symmetric configuration. After the transition, the locus continues along a curve with a gradually decreasing slope determined by $\omega_f \in [1.1480,  0.840)$. 

A key event in the curved segment of the locus is the transition of the forcing frequency from the primary resonance of the first mode of the system to the frequency of anti-resonance. The forcing frequency remains near the anti-resonance of the system along the second semi-linear segment of the locus of reciprocal response. Although the response is reciprocal in this segment, it is of little practical interest because the amplitudes of motion are very small (theoretically zero) at an anti-resonance; in other words, the measured amplitude response would be indistinguishable from noise. 

In summary, Fig.~\ref{fig:LocusNonReciprocity} shows two sets of parameter ranges for realizing reciprocal dynamics in the vicinity of the primary resonances of the system. For the second mode, reciprocity can be realized for a large range of $r$ and $\mu$.  For the first mode, in contrast, \textit{observable} reciprocity (not at an anti-resonance) can only be obtained for bounded values of $r$ and $\mu$, depending on the numerical tolerance used for $R\approx0$.
Increasing the forcing amplitude, $P$, can change the locus of reciprocal dynamics. The Appendix provides a more detailed account of this relation.

We see in Fig.~\ref{fig3} that for each set of parameters there are two points in the frequency response curve at which $N^F=N^B$. The number of such points depends, among other considerations, on where the nonlinear force appears in the system. For example, if nonlinearity appears only in the coupling force between the two oscillators, then phase nonreciprocity appears more easily (at a lower value of $P$) for the second mode~\cite{AliHooman}. 

The reciprocity invariance was broken by $\mu$ and restored with the help of $r$, both of which belong to the linear terms in Eq.~\eqN{eq:EOM}. However, the symmetry-breaking parameters may belong to nonlinear terms too. To show this, we replace $r$ with $\alpha$ as the second symmetry-breaking parameter~\cite{Ralpha}. Fig.~\ref{fig5} shows the frequency response curves of the system with $\mu$ and $\alpha$ chosen such that reciprocity is restored for the first mode (left column) and the second mode (right column). When restoring reciprocity for the first mode, the frequency response curves in panels (a1) and (a2) are very similar (though not identical) to the same panels in Fig.~\ref{fig3}. This is not the case for the second mode; cf. panels (b1) and (b2) in Figs.~\ref{fig3} and~\ref{fig5}. 
The output displacements of the restored reciprocal response in panels (a3) and (b3) of Fig.~\ref{fig5} are overall very similar to the corresponding results in Fig.~\ref{fig3} because $P=0.5$ in both scenarios. In effect, we are restoring the same reciprocal output displacement by balancing $\mu$ with $r$ in Fig.~\ref{fig3} and with $\alpha$ in Fig.~\ref{fig5}. We present and discuss the locus of reciprocal dynamics in the $(\mu,\alpha)$ plane in the Appendix.

In conclusion, we have studied the reciprocity invariance in coupled nonlinear systems with broken mirror symmetry. Although breaking the mirror symmetry is necessary for destroying the reciprocity invariance, we have shown that a second symmetry-breaking parameter can counteract the original asymmetry and restore the reciprocity invariance. We have computed the nonreciprocal phase shift of the asymmetric system and tuned two symmetry-breaking parameters such that the nonreciprocal phase shift vanishes. We have thereby reported on the existence of stable, steady-state nonlinear reciprocal dynamics in coupled systems with broken mirror symmetry. The restored reciprocal response occurs in the vicinity of the primary resonances of the system. We hope these findings contribute to the development of devices and materials with new modes of operation and functionalities for controlling and steering of waves. This may particularly prove relevant in applications that are developed for symmetric devices but are limited by the inherent symmetry-breaking imperfections caused during manufacturing.



\section*{Acknowledgments}
A.G. was supported by KIAS Individual Grant No. CG086101 at Korea Institute for Advanced Study. B.Y. acknowledges support from the Natural Science and Engineering Research Council of Canada through a Discovery Grant. 

\nocite{*}

\bibliographystyle{unsrt} 
\bibliography{GY_Reciprocity.bib}

\begin{thebibliography}{10}
\bibitem{modes}
The first and second modes in Eq.~(1) are also known as the in-phase and
  out-of-phase modes. We avoid this terminology to prevent possible confusion
  of the nonreciprocal phase shift, $\Delta\phi$, with the internal phase
  differences between degrees of freedom within each of the forward and
  backward configurations.
\bibitem{Ralpha}
There is only one possible symmetry-breaking parameter in Eq.~(1) that
  corresponds to nonlinear terms. If, for example, each oscillator included a
  quadratic term too, then reciprocity could have been broken and restored
  using the coefficients of the quadratic and cubic terms.
\end{thebibliography}

\appendix

\section*{Appendix}
\section{Prediction of the locus of reciprocal dynamics}

We can develop a first-order approximation of the locus of reciprocal dynamics, $R=0$, for finite but small amplitudes of motion (weak nonlinearity). The leading-order effect of nonlinearity can be incorporated by assuming the response to be harmonic. This assumption is verified numerically for the results presented in this work -- see the discussion on harmonicity following Fig.~\ref{fig2}. 

Using the complex notation, we present the steady-state response of the system as follows 
\begin{equation}
    \label{Hcomplex}
    x_k^{F,B}=\frac{1}{2}A_k^{F,B}\exp(i\omega_ft) +c.c.
\end{equation}
for $k=1,2$, where \textit{c.c.} denotes the complex conjugate term. Applying the harmonic approximation~\cite{NayfehMook}, Eq.~(\ref{eq:EOM}) is transformed to the following set of algebraic equations for the response amplitudes of the forward and backward configurations: 

\begin{align}
        \label{HB1}
        \left(1-\omega_f^2+2i\zeta\omega_f+(3/4)k_N|A_1^F|^2\right))A_1^F + k_c(A_1^F-A_2^F)&=P
        \\
        \label{HB2}
        \left(r-\mu\omega_f^2+2i\zeta\omega_f+(3/4)k_N\alpha|A_2^F|^2\right))A_2^F + k_c(A_2^F-A_1^F)&=0
        \\
        \label{HB3}
        \left(1-\omega_f^2+2i\zeta\omega_f+(3/4)k_N|A_1^B|^2\right))A_1^B + k_c(A_1^B-A_2^B)&=0
        \\
        \label{HB4}
        \left(r-\mu\omega_f^2+2i\zeta\omega_f+(3/4)k_N\alpha|A_2^B|^2\right))A_2^B + k_c(A_2^B-A_1^B)&=P
\end{align}

Phase nonreciprocity happens when $N^F=N^B$, which corresponds to $|A_2^F|=|A_1^B|$. When the nonreciprocal phase shift is additionally set to zero, $\Delta\phi=0$, then the periodic orbits of the forward and backward configurations become identical and $R=0$. In this case, we have $x_2^F(t)=x_1^B(t)$, which corresponds to $ A_2^F=A_1^B$. 

In our computations of the locus of reciprocal response, $R \approx 0$ for example in Fig.~\ref{fig:LocusNonReciprocity}, we have observed that the amplitudes of the input displacements are equal to each other; i.e. $A_1^F=A_2^B$. Using this observation, in addition to the condition that $A_2^F=A_1^B$, we can subtract \eqN{HB3} from \eqN{HB2} to obtain 
\begin{equation}
    \label{eqLocus}
    (r-1)-(\mu-1)\omega_f^2+(\alpha-1)|A_2^F|^2=0
\end{equation}
where $\omega_f$ is the value of forcing frequency along $R=0$. The same equation is obtained if \eqN{HB1} is subtracted from \eqN{HB4}. When only $r$ and $\mu$ are used as the symmetry-breaking parameters, then $\alpha=1$ and we retrieve Eq.~(3) in the main text. 

Eq.~(\ref{eqLocus}) does not provide the value of $\omega_f$ along $R=0$. However, we have confirmed numerically that the computed loci in Fig.~4 satisfy Eq.~(3). 

Finally, we note that it would have been possible to solve Eqs.~(\ref{HB1})--(\ref{HB4}) to obtain an approximation of the harmonic response of the system and perform all the relevant computations presented in this paper. The obvious shortcoming of this approach is that it is limited to the harmonic response regime; moreover, evaluation of the stability of the response would require further analysis. 

\section{Higher values of P}
\begin{figure}[b]
  	 \centering
    	 \includegraphics{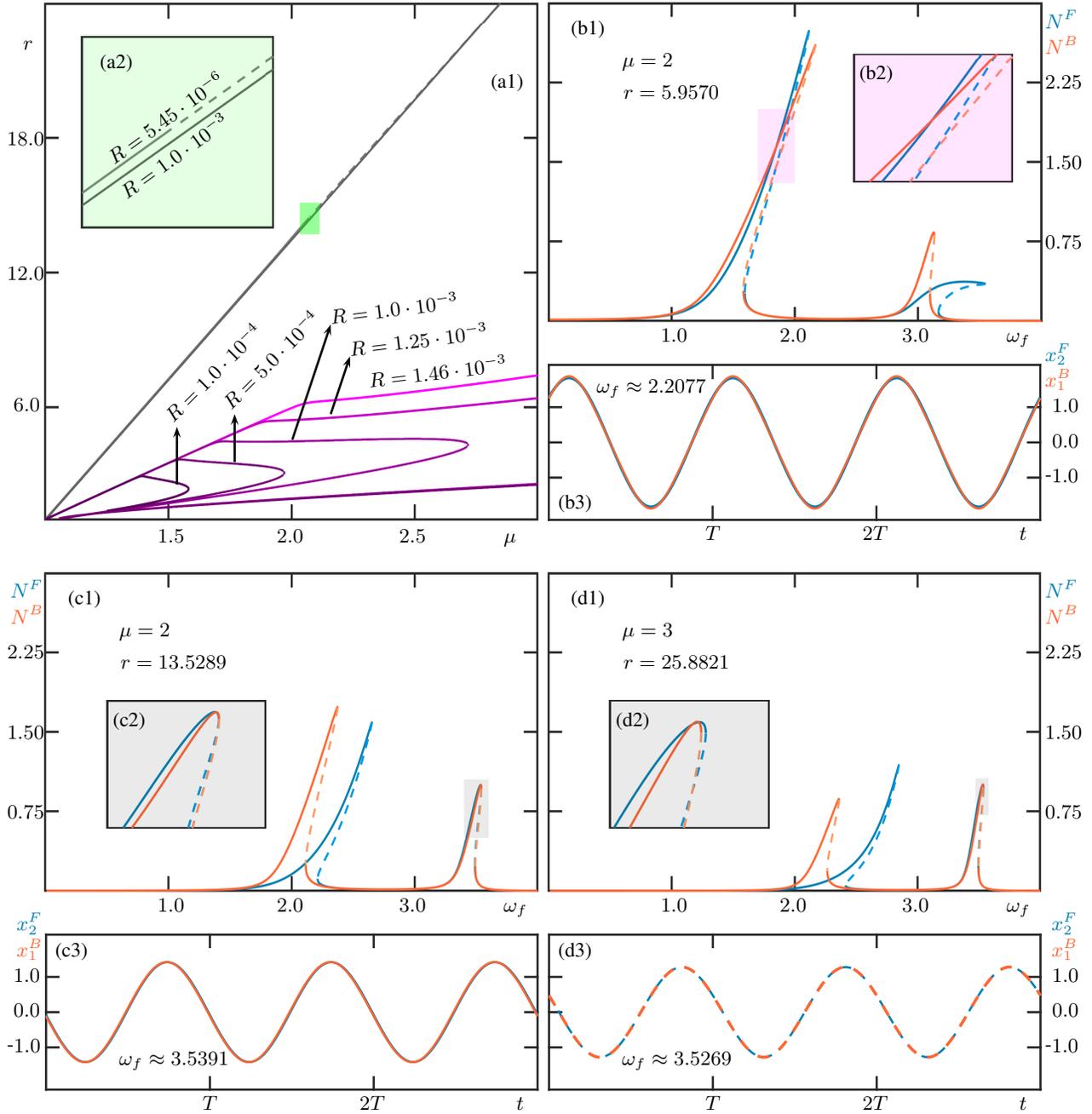}
   	 \caption{Loci of approximate reciprocity $R=\epsilon$ in the $(\mu,r)$-parameter plane for $P=1$, shown in panels~(a), and representative norms of the forward and backward output displacement at particular $(\mu,r)$ values along selected loci, shown in panels~(b) to (d). Panels~(2) show magnifications of the respective colored regions in their corresponding panels~(1). Panels~(b3), (c3) and (d3) show the corresponding reciprocity response of (b1), (c1) and (d1), respectively.  In panel~(1), purple lines correspond to the locus of reciprocity in the first resonance regime, where darker lines represent lower values of $\epsilon$, while grey lines correspond to the locus of reciprocity in the second resonance regime. In all panels, solid lines represent stable solutions and dash-lines unstable solutions. Other parameters in Eq.~\eqN{eq:EOM} which are not indicated in the figure are $(\zeta,k_c,k_N,P)=(0.05, 5, 1,1.0)$. } 
\label{fig:higherP}
\end{figure}

We discussed the locus of reciprocal dynamics ($R\approx0$) for $P=0.5$ in the main text; see Fig.~\ref{fig:LocusNonReciprocity}. Fig.~\ref{fig:higherP} shows the computed locus of reciprocity for $P=1$ in the $(\mu,r)$ plane. 

Panels (a1) and (a2) show two categories of parameter sets for approximate reciprocity, $R=\varepsilon$, for the first mode in purple and for the second mode in grey. 
The locus of the first mode is computed for five values of $\varepsilon$. The trend is very similar to those in Fig.~\ref{fig:LocusNonReciprocity}. Starting from the mirror-symmetric case, $(\mu,r)=(1,1)$, the locus follows a linear trend (constant $\omega_f$) up to a turning point. The locus then turns back towards the symmetric configuration and follows another linear trend. During this transition, the forcing frequency $\omega_f$ changes such that the point of approximate reciprocity is at anti-resonance (instead of being a primary resonance in the initial section of the curve). We recall that in the entire area swept by the purple curves, the value of $R$ is small; e.g. $R<1.46\cdot10^{-3}$ in the region in Fig.~\ref{fig:higherP}. 

Panels (b1)-(b3) show the frequency response curve for $(\mu,r)=(2,5.95701)$, where approximate reciprocity is obtained near $\omega_f=2.2077$ for the first mode. The response is stable at this point for both the forward and backward configurations. 

Returning to panel (a2), there is very little change in the locus of $R=\varepsilon$ for the second mode, except in terms of the stability of the output displacement. As we consider smaller values of $\varepsilon$, the reciprocity response becomes unstable past a point along the reciprocity curve. For the particular case of $\varepsilon = 5.45\cdot10^{-6}$, this change of stability occurs at $(\mu,r) \approx (2.07,14.48)$. Indeed, if one studies the response of backward and output displacements as a function of $\omega_f$ past this point, panel~(d1), one sees that the reciprocity intersection point occurs past a saddle-node bifurcation making the reciprocity response unstable for both displacements. This implies that at {\color{black}$(\mu,r) \approx (2.07,14.48)$} the reciprocity curve $R=5.45\cdot10^{-6}$ coincides with a saddle-node bifurcation, which marks the onset of instability for the responses.  Panels (c1)-(c3) show the frequency response curve and the time-domain response of the system for the forward and backward configurations for $(\mu,r)=(2,13.5289)$; that is, before the reciprocity response becomes unstable. Both the forward and backward configurations are stable near $\omega_f=3.5391$, where reciprocity is restored approximately with {\color{black}$R=5.45\cdot10^{-6}$}. In contrast, panels (d1-d3) show that both the forward and backward configurations are unstable near $\omega_f=3.5269$ for $(\mu,r)=(3,25.8821)$. Even though the output displacements at this frequency are very similar to those in panel (c3), the reciprocal response in panel (d3) is not stable and is therefore of little practical interest. However, notice in the inset~(d2) that the responses intersect in two additional points, which are stable but have a slightly higher value of R, which agrees with the existence of a stable reciprocity response at $R \approx 1.0\cdot 10^{-3}$ as shown in the inset~(a2).

\section{Locus of reciprocal dynamics in the $(\mu,\alpha)$ plane}

\begin{figure}[t]
  	 \centering
    	 \includegraphics{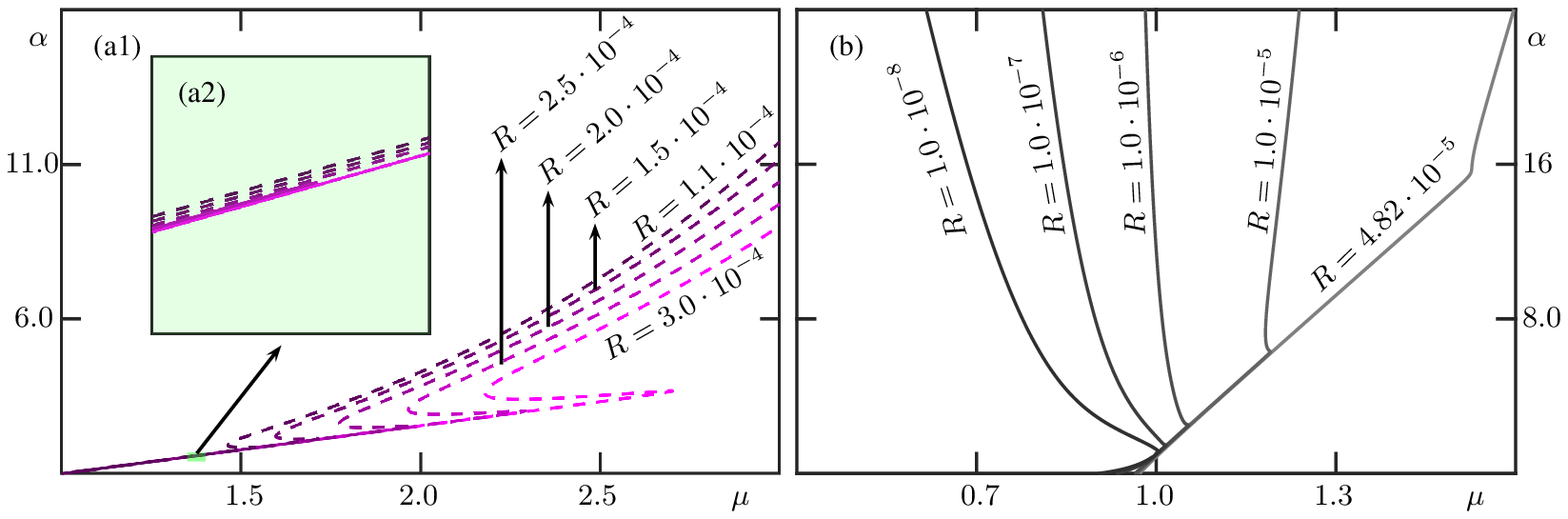}
   	 \caption{Loci of approximate reciprocity $R=\epsilon$ in the $(\mu,\alpha)$-parameter plane. Panel~(a) shows different loci of reciprocity in the first resonance regime, while panel~(b) shows select loci of reciprocity in the second resonance regime. Panel~(a1) shows magnifications of the respective and indicated colored region in panel~(a2). For all panels, darker lines represent lower values of $\epsilon$, solid lines represent stable solutions, and dash-lines unstable solutions. Other parameters in Eq.~\eqN{eq:EOM} which are not indicated in the figure are $(r,\zeta,k_c,k_N,P)=(1,0.05, 5, 1,0.5)$.} 
\label{fig7}
\end{figure}

The reciprocity invariance can be restored by tuning two symmetry-breaking parameters. Fig.~\ref{fig7} shows the locus of reciprocal dynamics $R\approx0$ for $P=0.5$ in the $(\mu,\alpha)$ plane. Panels (a1) and (a2) correspond to reciprocal dynamics at frequencies near the first mode, while panel (b) corresponds to the locus at frequencies near the second mode. In both cases, the locus passes through the mirror-symmetric configuration at $(\mu,\alpha)=(1,1)$, which is trivially reciprocal. From the symmetric configuration, both loci follow a linear trend (constant $\omega_f$) with the forcing frequency in the vicinity of a primary resonance; see Fig.~\ref{fig5}. There is a sharp turning point in each locus, which corresponds to the forcing frequency moving away from the primary resonance of the system towards the anti-resonance regime. The response remains stable along the locus of approximate reciprocity for the second mode, panel (b), for different values of $\varepsilon$. In contrast, panel (a2) shows that the stability of the response depends on the value of $\varepsilon$ for the first mode.

\end{document}